\documentclass{article}
\usepackage{epsfig}
\newread\epsffilein    
\newif\ifepsffileok    
\newif\ifepsfbbfound   
\newif\ifepsfverbose   
\newdimen\epsfxsize    
\newdimen\epsfysize    
\newdimen\epsftsize    
\newdimen\epsfrsize    
\newdimen\epsftmp      
\newdimen\pspoints     
\def\epsfbox#1{\global\def\epsfllx{72}\global\def\epsflly{72}%
   \global\def\epsfurx{540}\global\def\epsfury{720}%
   \def\lbracket{[}\def\testit{#1}\ifx\testit\lbracket
   \let\next=\epsfgetlitbb\else\let\next=\epsfnormal\fi\next{#1}}%
\def\epsfgetlitbb#1#2 #3 #4 #5]#6{\epsfgrab #2 #3 #4 #5 .\\%
   \epsfsetgraph{#6}}%
\def\epsfnormal#1{\epsfgetbb{#1}\epsfsetgraph{#1}}%
\def\epsfgetbb#1{%
\openin\epsffilein=#1
\ifeof\epsffilein\errmessage{I couldn't open #1, will ignore it}\else
   {\epsffileoktrue \chardef\other=12
    \def\do##1{\catcode`##1=\other}\dospecials \catcode`\ =10
    \loop
       \read\epsffilein to \epsffileline
       \ifeof\epsffilein\epsffileokfalse\else
          \expandafter\epsfaux\epsffileline:. \\%
       \fi
   \ifepsffileok\repeat
   \ifepsfbbfound\else
    \ifepsfverbose\message{No bounding box comment in #1; using defaults}\fi\fi
   }\closein\epsffilein\fi}%
\def\epsfsetgraph#1{%
   \epsfrsize=\epsfury\pspoints
   \advance\epsfrsize by-\epsflly\pspoints
   \epsftsize=\epsfurx\pspoints
   \advance\epsftsize by-\epsfllx\pspoints
   \epsfsize\epsftsize\epsfrsize
   \ifnum\epsfxsize=0 \ifnum\epsfysize=0
      \epsfxsize=\epsftsize \epsfysize=\epsfrsize
     \else\epsftmp=\epsftsize \divide\epsftmp\epsfrsize
       \epsfxsize=\epsfysize \multiply\epsfxsize\epsftmp
       \multiply\epsftmp\epsfrsize \advance\epsftsize-\epsftmp
       \epsftmp=\epsfysize
       \loop \advance\epsftsize\epsftsize \divide\epsftmp 2
       \ifnum\epsftmp>0
          \ifnum\epsftsize<\epsfrsize\else
             \advance\epsftsize-\epsfrsize \advance\epsfxsize\epsftmp \fi
       \repeat
     \fi
   \else\epsftmp=\epsfrsize \divide\epsftmp\epsftsize
     \epsfysize=\epsfxsize \multiply\epsfysize\epsftmp
     \multiply\epsftmp\epsftsize \advance\epsfrsize-\epsftmp
     \epsftmp=\epsfxsize
     \loop \advance\epsfrsize\epsfrsize \divide\epsftmp 2
     \ifnum\epsftmp>0
        \ifnum\epsfrsize<\epsftsize\else
           \advance\epsfrsize-\epsftsize \advance\epsfysize\epsftmp \fi
     \repeat
   \fi
   \ifepsfverbose\message{#1: width=\the\epsfxsize, height=\the\epsfysize}\fi
   \epsftmp=10\epsfxsize \divide\epsftmp\pspoints
   \vbox to\epsfysize{\vfil\hbox to\epsfxsize{%
      \special{illustration #1 scaled \number\epsfscale}
      \hfil}}%
\epsfxsize=0pt\epsfysize=0pt\epsfscale=1000 }%

{\catcode`\%=12 \global\let\epsfpercent=
\long\def\epsfaux#1#2:#3\\{\ifx#1\epsfpercent
   \def\testit{#2}\ifx\testit\epsfbblit
      \epsfgrab #3 . . . \\%
      \epsffileokfalse
      \global\epsfbbfoundtrue
   \fi\else\ifx#1\par\else\epsffileokfalse\fi\fi}%
\def\epsfgrab #1 #2 #3 #4 #5\\{%
   \global\def\epsfllx{#1}\ifx\epsfllx\empty
      \epsfgrab #2 #3 #4 #5 .\\\else
   \global\def\epsflly{#2}%
   \global\def\epsfurx{#3}\global\def\epsfury{#4}\fi}%
\newcount\epsfscale    
\newdimen\epsftmpp     
\newdimen\epsftmppp    
\newdimen\epsfM        
\newdimen\sppoints     
\epsfscale=1000        
\sppoints=1000sp       
\epsfM=1000\sppoints
\def\computescale#1#2{%
  \epsftmpp=#1 \epsftmppp=#2
  \epsftmp=\epsftmpp \divide\epsftmp\epsftmppp  
  \epsfscale=\epsfM \multiply\epsfscale\epsftmp 
  \multiply\epsftmp\epsftmppp                   
  \advance\epsftmpp-\epsftmp                    
  \epsftmp=\epsfM                               
  \loop \advance\epsftmpp\epsftmpp              
    \divide\epsftmp 2                           
    \ifnum\epsftmp>0
      \ifnum\epsftmpp<\epsftmppp\else           
        \advance\epsftmpp-\epsftmppp            
        \advance\epsfscale\epsftmp \fi          
  \repeat
  \divide\epsfscale\sppoints}
\def\epsfsize#1#2{%
  \ifnum\epsfscale=1000
    \ifnum\epsfxsize=0
      \ifnum\epsfysize=0
      \else \computescale{\epsfysize}{#2}
      \fi
    \else \computescale{\epsfxsize}{#1}
    \fi
  \else
    \epsfxsize=#1
    \divide\epsfxsize by 1000 \multiply\epsfxsize by \epsfscale
  \fi}

\begin{document}

\title{Matter Outflows from AGN:
A Unifying Model}

\author{E.Y. Vilkoviskij$^a$, R.V.E. Lovelace$^b$, L.A. Pavlova$^a$,\\
M.M. Romanova$^c$,
and S.N. Yefinov$^a$}

\maketitle

\begin{abstract}
We discuss a self-consistent unified
model of the matter outflows from AGNs
based on a theoretical approach and
involving data on AGN evolution and
structure. The model includes a unified
geometry, two-phase gas dynamics,
radiation transfer, and absorption
spectrum calculations in the UV and X-ray
bands. We briefly discuss several
questions about the mass sources of the
flows, the covering factors, and the
stability of the narrow absorption
details.

\medskip
\noindent(a){Fesenkov Astrophysical Institute of the
National Academy of Science of Kazakhstan, Almaty,
Kazakhstan, 480020 Almaty, Kazakhstan; vilk@aphi.kz}

\noindent(b)~{Departments of Astronomy and
Applied and Engineering Physics, Cornell University,
Ithaca, NY 14853; RVL1@cornell.edu}

\noindent(c)~
{Department of Astronomy, Cornell University, Ithaca, NY
14853-6801;  romanova@astro.cornell.edu}

\end{abstract}
\noindent{Subject headings: galaxies: active -- quasars -- galaxies: nuclei --
accretion disks -- outflows}

\section{Introduction}
 There are two apparent manifestations of matter
outflows from active galactic nuclei (AGN):
relativistic jets observed mostly in radio-bands,
and gas outflows visible in absorption in the UV
and X-ray bands. The only hint of a connection of
the two outflows is that the broad absorption lines
are mostly observed among the radio-quiet objects,
but the physical reason of this anti-correlation is
unknown.  Regarding the jets, there is strong
support for the electromagnetic nature of the
accelerating and collimating forces
(Bisnovatyi-Kogan \& Lovelace 2001; Blandford \&
Znajek 1977; Lynden-Bell 1996; Lovelace 1976;
Lovelace \& Romanova 2003).
   However, there is no
common approach to the theory of absorbing
outflows. Here, we discuss the outflows of the
second type which were first observed in the broad
absorption line quasars (BALQSO) and now are widely
accepted as the physical reason for blue-shifted
absorptions in both quasars and Seyfert
galaxies.

The BALQSOs were discovered at the end of the
1960's (Lynds 1967; Burbidge 1970) and have
attracted increasing attention from the beginning
of 1980's (Weymann, Carswell, \& Smith 1981;
Turnshek 1987; Weymann et al. 1991).
  The intrinsic
UV absorptions were observable in the optical band
due to the large redshifts of the objects ($z \sim 2$).
Around 2000, rapid development of space telescopes
brought many new results, including precise
observations of absorption in the UV and X-ray
bands of the AGN spectra (Crenshaw 1997; Reynolds
1997; George et al. 1998a; Kriss et al. 1995; Kriss
et al. 1996; Crenshaw et al. 1998; Kaastra et a.
2000; Kaspi et al. 2000). In the new century the
data regarding absorbing outflows from AGN grows
quickly. Chandra and XNN-Newton have detected a
wealth of absorption features (Komossa and Hasinger
2002; Netzer et al. 2003). But still there is no
commonly accepted theory (or even model), which
explains the physical properties, geometry, and
dynamics of the absorbing flows. The X-ray spectra
obtained with the new generation of telescopes
appear to be more complex then those calculated
with any previously considered models. As a result
some emphasis has been given in recent years to
empirical models and interpretations of observed
properties of absorption features.

Here, we trace some the ideas and attempts
at interpretation of the absorbing spectra.

The earliest theoretical models contained
some general ideas, including the possible role of
drag forces and cosmic rays in the dynamics of a
two-phase medium (Weymann et al. 1982; Begelman,
de Kool, \& Sikora 1991). Most of the following
dynamical models explored mainly the radiation
pressure force by analogy with the theory of hot
star's wind (i.e., Lucy \& Solomon 1970; Castor,
Abbot \& Klein 1975). Arav and  Begelman (1994)
considered arbitrary radial outflows, while Murray
et al. (1995), Murray and Chang (1995), (1998),
and Proga, Stone and Kallmann (2000) considered
the flows above the surface of an accretion disk.
Though the models predict UV and X-band
absorptions, many questions about the structure
and physics of the flow remains unanswered. The
limitations of the dynamical models led to the
development of purely empirical models (Elvis
2000; Ganguly et al. 2001).  These were
constructed to explain the wealth of observational
data qualitatively without dynamical
considerations. Unfortunately, all the various
models do not unambiguously solve the main
problems of AGN outflow physics: Where and why are
the absorbing outflows generated in AGNs, and what
determines the structure and dynamics of the
outflows? Most of the models were built especially
for interpretation of the observational data on
absorption details without accounting for the full
picture of the AGN structure and evolution.

In the present paper we discuss the approach
to physics of the AGN outflows, referring to the
well-established AGN unification and evolution
model, as well as to important observational data.

The structure of the paper is as follows:
In \S 2 we discuss relations of the
evolution, unification, and mass outflow problems
and present the geometrical structure of our
model.
  In \S 3 we recall the main points of
our model and discuss some peculiarities of a
two-phase medium.
  In \S 4 sample results of
the model calculation of AGN spectra in the UV and
X-ray bands are presented.  Section 5 gives the
main conclusions.

\section{Interacting Subsystems of AGNs,
Evolution and Unification Scheme}

Our theoretical approach to the problem of
matter outflow from AGN was presented in a
series of papers by Vilkoviskij and Nosov
(1994), Vilkoviskij, Karpova and Nosov (1996),
Vilkoviskij and Karpova (1996), and in complete
form by Vilkoviskij et al. 1999 (hereafter
V99). Here, we recall the main points of this
approach and discuss them in the light of new
observational data. In V99 we presented the
theory of BALQSO outflows based on the
``interacting subsystems approach'', which
supposes that the AGN consists of three main
physically distinct, but strongly interacting
subsystems: the central super-massive black
hole (MBH); the compact stellar cluster (CSC),
and the gas subsystem. The last subsystem
usually includes the accretion disk (AD), the
obscuring torus (OT), and the hot gas (HG)
flowing within the hole of the torus. In V99 we
have introduced the CSC and HG as ``hidden''
subsystems in the sense that there was
insufficient observational data to constrain
them at that time.

Now such data starts to appear, but still its
interpretation remains controversial. The observational
signs of high-ionization species of many elements were
provided by the present generation X-ray missions
XMM-Newton and Chandra, which revealed new features in
the spectra of AGN. Both the emission lines from Fe XXV
and Fe XXVI ions (Bianchi et al. 2004; Matt et al.
2004b) and those lines in absorption (Kaspi et al.2002;
Reeves et al. 2003; Matt et al. 2004a; Reeves et al.
2004) were observed in recent years. The interpretation
of the data (Bianchi \& Matt 2002; Bianchi et al. 2004,
Bianchi et al. 2005) is in agreement with the
orientation-depended unification model with a wide
distribution of temperatures of the outflowing matter,
up to $T\sim 105 - 107$ K. Nevertheless, the interpretation
is model-dependent.  It depends on the assumed
structure of the gas and so cannot be considered as a
direct proof of the high-temperature gas. The analysis
of the X-ray emission lines in NAGC 1068 (Ogle et al.
2000, 2003) lead the authors to the conclusion that the
line emission dominated by warm ($T_e\sim 0.8\times105$ K),
photo-ionized gas, moving with outward velocity less
then $1000$ km/s, possibly coincident with the narrow
emission line clouds seen in the optical band. The
presence of collision-ionized hot gas with $T\sim 107$K is
restricted by the emission measure
$EM < 3 \times10^{63}{\rm cm}^{-3}$, where
$EM=V n2$.  However, the authors note that pressure
confinement of the observed warm clouds requires the
presence of a hot component with $T_e \sim 107$ K (Krolick,
McKee, \& Tarter 1981). So, our conclusion is that,
though there are no direct evidences of the hot gas
($T \sim 107 - 108$ K) in the outflow, its presence is
possible and it is supported by confinement and
dynamical  arguments discussed below.

The second subsystem postulated in V99 was the
compact, massive stellar cluster (CSC).  Its mass
supposed to be larger than or of the order of the
central black hole mass, and its size is around several
parsecs. Later, we considered the physical properties
and evolution of such CSC due mainly to its interaction
with the accretion disk (Vilkoviskij and Czerny 2002;
hereafter VC02). We showed that the orbits of stars in
the central part of the CSC tend to concentrate in the
plane of the accretion disk.  However, due to star-star
interactions the orbits do not coincide with the
accretion disk plane as found for the case of a single
star (Syer, Clarke, \& Rees 1991). We concluded that the
mass loss from the stars crossing the accretion disk
and star-star collisions could supply both the hot-gas
outflow from the disk and the matter inflow to the MBH.
  Now the first evidence of the CSC in AGNs have been
provided by recent observations of the AGNs structure
with $0.1''$ resolution scales (Davies, Tacconi and Genzel
2004 a, b). Also, an indication of the reality of the
CSC in AGNs can be seen in ``relict'' clusters
frequently observed in the centers of ``quiet''
galaxies containing SMBH (Kormendy 2001) and in the
stellar clusters in the centers of late-type galaxies
(Boeker et al 2004).

There is substantial evidence that AGN evolution
is driven mainly by intergalactic interactions and
merging (e.g., Sanders et al. 1988; Menci et al. 2003,
and references therein).  Evidence for merged
systems includes the counter-rotating galaxies
(Rubin, Graham, \& Kenney 1992;  Kuznetsov et al. 1999).
  Every merging event leads to a
new ``duty cycle'' of the AGN activity, accompanied by
a starburst event in the earlier stages. The first
phase of the cycle starts with the creation a new CSC by
a powerful starburst in the massive dense gas cocoon
around the MBH.  This corresponds to the brightest IRAS
galaxies.

The gas and stars in the center of the cocoon
have a definite specific angular momentum.  This
angular momentum determines the orientation of the
new accretion disk and the CSC around the central
MBH.  Hence, the symmetry axis of the AGN is not in
general coplanar with the galactic disk. At the
end of this phase, gas outflows and jets
perpendicular to the disk produce polar holes in
the cocoon along the symmetry axis, transforming
it to a typical ``obscuring torus'' (OT). Then the
bright AGN phase begins and it lasts about $107-108$
years.
  In this bright phase of the AGN duty cycle,
the masses of the CSC and OT gradually decrease,
leading to the third phase with diminishing
activity corresponding to weak Seyfert nuclei and
LINERs. The corresponding evolution sequence of
OTs during the duty cycle is shown schematically
in Figure 1 (a, b, c). It is similar to the
evolution scheme derived from observations of the
infrared (IR) spectra of AGNs by Haas et al.
(2003).

This outlined picture of AGN evolution is compatible
with the standard geometrical unification model of
Antonucci (1993). The beautiful unification idea argued
for the simplicity and similarity of all AGNs.  But it
demanded the introduction of a new important element of
AGN structure, the obscuring torus (OT) (Antonucci and
Miller 1985, Krolik and Begelman 1988). Nevertheless,
we are still far from understanding the physical nature
of OT. Obviously, the physics of the OT is connected
with the existence of the compact stellar cluster
produced by a starburst in the center of AGN (Wada,
Norman, \& Colin 2001, 2002). From the above-sketched
scheme of AGN evolution, the axial hole in the torus
can naturally be produced by of the above-mentioned
polar outflows of hot gas through the dusty, rotating
CSC. Taking into account that the hot-gas outflow
inevitably drags some cold matter from the internal
surface of the OT, we conclude that the unification
scheme must also include the outflow structure.

\section{Unified Model of the Absorbing Mass
Outflows}

From the evolution scheme presented above a unified
model follows.  It includes both the ``classical''
AGN1-AGN2 unification and the absorbing outflow models.
The model predicts the appearance of the spectral
absorption features, the broad absorption lines in the
UV, and the absorptions both in lines and continuum in
the X-ray band. A sketch of the model is shown in
Figure 2.  The hot gas outflow, shown with radial
arrows, produces a two-phase medium in the ``transition
layer'' at the internal surface of the obscuring torus
(clouds are shown with ovals and dots).  The arrow
marked with the ``R'' indicates reflected
radiation.

We call the model a ``Unified Outflow Model'' in the
sense that it is based on the standard Unification Model
(Antonucci 1993) and derived the outflow properties
from the unifying geometry. Taking into account this
geometry, one has to consider the internal surface of
the obscuring torus illuminated by the radiation flux
from the central part of AGN. According to analyzes by
Krolik, McKee, and Tarter (1981), Kwan and Krolik (1981),
Krolick and Kriss (1995, 2001), a two-phase medium is
produced at the internal surface of the OT.

We consider the dynamics of a two-phase medium
consisting of cold clouds imbedded in a hot gas.
Analysis of the thermal balance of the two-phase medium
in the central region of AGNs shows that the
temperature of the hot gas is of the order of $107$
to $108$ K
(Kwan \& Krolik 1981; Fabian et al. 1986;
Sazonov, Ostriker, \& Sunyaev 2004).
Consequently, the electron heat conductivity is
sufficiently high that the hot gas can be treated as
isothermal to a first approximation.   Thus the hot gas
wind equation has a form close to Parker's equation,
namely,
$$
\left(1-{a2 \over v2}\right) v {dv \over dr}=
{2 a2 \over r} - {G M(r) \over r2} =g_{\rm drag}
+g_{\rm rad}~,
\eqno(1)
$$
where $v$ is the velocity of the hot gas,
$a = \sqrt{k T/m_p}$
is the  isothermal sound speed of the hot
gas, and $r$ is the radial distance.
  The terms on the right-hand side of
the equation are the accelerations due to the hot gas
pressure gradient, to gravity, to the drag force from
the cool clouds, and to the radiation pressure
(due to the Compton scattering mainly in the hot gas).

The notable difference of equation (1) from the Parker
equation is that  $M(r)$ is the total mass
inside the radius $r$, which is sum
of the black hole mass and the
distributed mass of the compact stellar cluster (CSC).
Because of this the solution of the equation (1) can
contain three critical points instead of one presented
in the Parker solar wind equation (Vilkoviskij \&
Karpova 1996). In this case the position of the
trans-sonic point depends on the parameters of the
equation; the most important parameter is the relation
of the masses of the black hole
and the CSC (Vilkoviskij et al. 1999).

      The motion of the cold clouds imbedded in the hot gas is
determined mainly by the radiation pressure force $F_{\rm rad}$,
the drag force $F_{\rm drag}$ by the hot gas, and the
gravitational force $F_g$. Thus the equation of motion of a
single cloud is
$$
m_{cl}{dV \over dt}= F_{\rm rad}+F_{\rm drag} + F_g~,
\eqno(2)
$$
where $m_{cl}$ is the cloud mass and $V$ is the cloud
velocity, and
$F_{\rm rad}=F_{\rm lin}+F_{cont}$ is the sum of
the forces from line-scattering processes and continuum
absorption due to photoionization, Compton scattering,
and dust absorption.

    The drag force depends on the square of the velocity
difference between the cloud and the hot gas,
$$
 F_{\rm drag}=\rho_{hg}S_{cl}\big[v(r) - V(r)\big] \big| v(r) - V(r)\big|~,
\eqno(3)
$$
where $\rho_{hg}$ is the hot gas density, and
$S_{cl} =\pi\{ m_{cl}/
[4/3¼\mu_{hg}T_{hg}\mu_{cl}/(T_{cl}\mu_{hg})]\}^{2/3}$ is the
projected surface area of a spherical cloud.
  The pressure balance
condition is $(\rho_{cl} /\mu_{cl} )T_{cl} =
(\rho_{hg} /\mu_{hg} )T_{hg}$, where
$\rho_{cl},~ \mu_{cl}$, $T_{cl}$,
and $\rho_{hg}$, $\mu_{hg}$, $T_{hg}$  are the
density, molecular weight, and temperature of the cloud
and the hot gas, respectively.

Calculation of absorption spectra includes two tasks:
calculation of the absorption in a single cloud and
calculation of the absorption in the cloudy medium.
 The calculation for a cloud can be divided
into two steps.
   First we solve for the ionization
radiation transfer in the cloud. This allows the
calculation of the change of the intensity of the
radiation with specified energy, $I(E)dE$, through the
cloud  and the distribution of the density of ions of
the most abundant elements in the cloud, as well as the
total column density of the ions $N_i$.
  These quantities
permit us to calculate the optical depth of the cloud
both in spectral lines and in the continuum.
  From this we can calculate the radiation transfer in the cloudy
medium.

  Calculation of the radiation transfer in a cloudy medium
can be treated in the following way.
   We assume that the
radiation transfer in the spectral lines is determined
by resonance scattering on the ions.
Also, we take into account the first scattering only,
ignoring multiple scattering. This is possible
due to non-spherical outflow of the cold clouds (see
Fig.1).

We determine the optical depth in the $j-$th line
center  $\tau_j=(\pi  e2/m_ec)(\int d\ell N_i)f_j/\Delta \nu_{j}$.
Here, $e$ and $m_e$ are the
electron charge and mass, $\int d\ell N_j$ is the column density of
the ions in the cloud (which scatter photons of the
frequency $\nu_j$),  $f_j$ is the oscillator strength of the
transition, and $\Delta \nu_{j}=\nu_j v_T/c$  is the turbulent Doppler
width of the line with $v_T$  the turbulent velocity and $c$ the speed
of light.
  We suppose that the absorption line
has a normalized Gauss profile $\varphi(\nu-\nu_j)$.
  Each cloud
absorbs a fraction $1-\exp(-\tau_j)$  of the radiation flux
in the line center.
   Also, the clouds can shield each
other. Then, the differential equation for the
radiation flux spectral density $\Phi(\nu)$ has the form
$$
{d\Phi(\nu) \over dr}=-\Phi(\nu)N_{cl} S_{cl}
\bigg[1-\exp\big(-\sum_j \varphi(\nu-\nu_j)\tau_j\big)\bigg]~
\eqno(4)
$$
where $N_{cl}$ is the number of  clouds per unit
volume, $S_{cl}$ is the cross-section of a cloud.
The probability of crossing a cloud
in a distance $dr$ is $dr S_{cl}N_{cl}$.
  The frequency which is scattered in the line center is
the Doppler-shifted one relative to the spectrum of the
central source, $\nu_j = \nu_{j0}[(1+V/c)/(1-V/c)]^{1/2}
\approx \nu_{j0}(1+V/c)$,
where $V$ is the velocity of the cloud, $\nu_{j0}$ is
the rest frame frequency in the line center.
   Thus
equation (4) takes into account both the filling factor
and the velocity of the clouds.

     A numerical code for solving the full system of
equations described above has been developed by Vilkoviskij
et al. (1999).
  The solutions for the gas dynamics and resulting spectra
were obtained under the following conditions:
 (i) The incoming AGN spectrum consists of Planck
and power-low parts and includes the broad emission
lines as well.
 (ii) The hot gas flow is time independent.
The mass flux of the cold clouds is determined with a
function which typically starts from a very small value
at  small radii, has a maximum at the distance of the
obscuring torus, and then decreases.
  That is, we
suppose the clouds are generated due to interaction of
the hot gas with the internal surface of the OT.
 (iii) At every step of the solution we calculate
the cloud's ionization and dynamics, and change of the
radiation flux spectral intensity due to both line and
continuum absorption. We take into account $374$ lines,
$147$ ion species, and $12$ of the most abundant elements.

Figure 2 shows the geometrical essence
of the unification only.
   The unification model for an
actual AGN has to include the evolutionary stage of the
AGN according to the Figure 1 as well as the type of the
host galaxy.  In particularly, the evolutionary stage
is definitely related to the nature of the MgII BALQSO
(the low-ionization BALQSO).
  There is evidence for
both the influence of
the orientation and for the influence of the starburst
absorption in many of objects (Hines et al. 2001).

The next question related to the models of actual
AGNs is the source of the mass outflows  at
different temperatures (the cold and warm absorbers).
For example, what are the mass sources, the places, and
the driving forces of the hot gas outflows?

In our view the most probable mass source of the
hot gas is the hot corona above the accretion disk
(Czerny \& Lehto 1997; Czerny, Schvarzenberg-Czerny \&
Loska 1999; Kawaguchi, Shimura, \& Mineshige 2001),
which creates a hot gas wind.
  Magnetohydrodynamical
driving from the surface of the accretion disk is also
a likely source of the hot gas outflows (Romanova et al.
2005; Ustyugova et al. 2006).
   The hot gas outflow is
assumed sufficiently strong that it acts to
entrain cold clouds of the OT into the flow.
   The large clouds fragment into  smaller ones which are
heated and gradually evaporated, creating cold and warm
absorbers (Reynolds 1997; George et al. 1998b).
 Of course, each cold cloud is surrounded with a warm
envelope due to both heat conduction and to its motion
through the hot gas in AGN radiation field.
Consequently, a multi-phase outflow is produced along a
``conical surface'' with a solid angle of the outflow
corresponding to the relative fraction of
sources which of AGN1's.

Krolik and Kriss (1995, 2001) have argued that
specific conditions at the internal surface of OT lead to
the creation of a multi-phase medium due to the
evaporation of cold matter of the OT at the isobaric
surface ($P=$const).
  Blustin et al. (2005) investigated the
properties of the warm absorbers in $17$ AGNs which produce
photoelectric absorption features in the spectra of
Seyfert galaxies and BALQSOs. These are associated in most
cases with the broad absorption lines in the UV band
(Crenshaw et al. 1999).
   Analysis of the properties of the
absorbers shows that their distances from the central
engine of Seyfert galaxies are typically close to the
sizes of the OT, which supports the ``multiphase torus
wind'' model.

The first arguments supporting the ``unified'' outflow
picture were based on the results of spectropolarimetry of
BALQSOs (Hines and Wills 1995; Ogle 1997; Ogle et al.,
1999).
   These results showed very strong polarizations (up
to $\sim 10\%$) in the broad absorption lines.
  The deeper the
absorption ``troughs'' the stronger  the polarization.
   This was interpreted as a mixture of the light reflected
from the internal surface of the OT (the arrow marked ``R''
in Fig.2).
   The inclination of the line of sight from the
torus axis defines the transition from AGN1 to the AGN2
types.
   A good tracer of the inclination angle is the
absorption in the X-ray band (Véron-Cetty \& Véron 2000).
The objects of type $1.5$ (such as NGC 4151 and
NGC 3516) may be examples of intermediate inclination
angles (Kriss et al., 1992).

The present unification model does not exclude the
accretion disk as a source of the obscuring matter.
In particular, some properties of the models suggested by
Murray et al. (1995), Proga, Stone, and Kallmann (2000),
as well as empirical models (Elvis 2000; Ganguly et al.
2001) can provide some matter observable in absorption.
   This can happen if gravitational instability or magnetic
fields strongly increase the thickness of the outer part
of the accretion disk.
  The ``inflating'' effect of an ordered magnetic
field on the OT has been studied by Lovelace, Romanova,
and Biermann (1998).
  Further, the warping of the disk at large
distances may be important.

\section{On the Dynamics of the Two-Phase Medium and
Line-Locking Interpretation}

Here, we briefly discuss several problems of
interpretation of the absorption spectra in the UV and
X-ray bands, which  were noted by Weymann
in his review of the conference devoted to the
mass-outflow problem (Weymann 2002).
The key problems are:
(i) The ``local covering factor.''  This is the fraction of
the continuum radiation source shadowed by the absorbing
gas at a defined velocity.
   It has no commonly
accepted interpretation (Barlow et al. 1997; Arav et al.
2002, 2003;  deKool et al. 2002);
(ii) Differences of the column densities derived from X-ray
and UV absorptions. Usually the latter seems to be much
less (Arav N.et al., 2003);
  (iii) Stability of the narrow details in the structured
absorption profiles of some objects, both in luminous
BALQSO (Foltz et al. 1986) and in low luminosity AGNs
(Weymann et al. 1997).  The brightest instances are
Q1303+308 and NGC 4151.

According to the main points of the unified model
of matter outflow described above, one has to consider
the dynamics of the hot-gas outflow and the dynamics of
the cold clouds embedded in the hot gas.  This was taken
into account in our model (V99). The main points of our
numerical model are:

1) Spherical ``cold''  clouds ($T_{cl}\sim 104$ K) move in
the hot gas ($T_{hg} \sim 106 - 108$ K) in pressure equilibrium.
The acceleration of the clouds is due to the radiation
pressure both in lines an continuum.  Also, we
account for the drag force
due to the different speed of the hot gas and gravity.

2) The hot gas flow is driven radially by the
pressure gradient acting against  gravity. The gravity
is determined by both the central black hole and the compact
stellar cluster (CSC).

3) The radiation transfer is treated in the cloudy
medium where only first scatterings of photons are
accounted for.

We used the equation for the change of the
radiation flow $\Phi(\nu)$ due to absorption of radiation in
spectral lines in the form equation (4). The equation
for continuum photoelectric absorption is:
$$
 {d\Phi(\nu) \over d\nu}=
-\Phi(\nu)N_{cl}S_{cl}\big\{1-\exp[-\tau(\nu_D)]\big\}~,
\eqno(5)
$$
(V99) where $\nu_D = \nu(1+V_{cl}/c)$.

    Here, we consider the reason for the differences
in the estimates of the column densities resulting from
the absorption features in the UV and X-ray bands.
  We first estimate the dependence of the absorption on the
acceleration of the clouds.
  The Doppler width of a
spectral line is $\Delta \lambda= \lambda_0 V_T /c$,
where $V_T$ is the
characteristic turbulent velocity in the cloud.
   Letting  $a=dV/dr$, we find that the velocity
of a cloud grows from $V$ to $V+ V_T$ in a distance
$\Delta r=V_T/(dV/dr)$.
   The average number of clouds
shadowing the source of the continuum radiation (from
the accretion disk) is $N \sim n_{cl}\Delta r S_d$,
where $n_{cl}$ is the
number of clouds per unit volume
and  $S_d$ is the surface
area of the disk.
  Thus the ``local covering factor'' by
clouds with cross-section $S_{cl}$
in a slab of thickness $\Delta r$ at
velocity $V(r)$ is $F(V)
\sim N(S_{cl}/S_d) \sim n_{cl} S_{cl} V_T/(dV/dr)$,
where $\Delta r$ is ``Sobolev's length''.
  Obviously, the
covering factor at a given velocity can be small when the
velocity gradient is large.
   Accordingly, the apparent
absorption in a line will be small in this case, even if
the line absorption depth is large in every cloud. As
shown by Kwan (1990), the apparent opacity depends on
the covering factor $F(V)$ and the single cloud opacity
$\tau_{cl}$ as
$$
\tau(V)= -\ln[\Phi(V)/\Phi_0]=  \sqrt{\pi}F(V) q(\tau_{cl})~,
\eqno(6)
$$
where
$$
     q(\tau_{cl}) \sim \tau_{cl}(1-0.28\tau_{cl})~,
\quad{\rm for}\quad
\tau_{cl} \ll 1~,\quad \quad \quad
$$
$$
q(\tau_{cl}) \sim [1.46 \ln(\tau_{cl}+0.43)]^{1/2}~,
\quad {\rm for}\quad
\tau_{cl} \gg 1~.
$$
The situation is quite different in the case of
continuum absorption, because the half-widths of the
absorption bands (in the case of photoelectric
absorption) are much wider then the turbulent widths
of the absorption lines.
  This could, in principal,
explain the different estimations of the column
densities from UV line absorptions and the X-ray
continuum absorption.

\section{Line-Locking Interpretation}

The dependence $\tau(V)$ of equation (6) on the velocity
gradient $dV/dr$ is similar to Sobolev's optical depth
$\tau(V)$ in a moving gas.
  The difference is that in the
system of clouds every cloud can be considered as a
``macro-atom'' with multiple lines.
   The considered
dependence of the covering factor on the cloud
acceleration is also related to the so-called
``line-locking effect'' (V99) due to the nonlinear
relation of the radiation pressure force on the
acceleration.
  Let us consider this effect qualitatively.
The acceleration of clouds by radiation pressure is
determined with both the line and the continuum absorption, but
in the small clouds (with line opacity $\tau_{cl}<1$), the
acceleration by lines dominates ($T_{cl} <3\times 104$ K).
  The radiation acceleration by lines is typically larger than
the acceleration of the hot gas.
   As a consequence, the
clouds ``out run'' the hot gas (velocity $v_{hg}$ ).
 That is, they move with velocity
$V> v_{hg}$, which gives rise to
the drag-force $F_{\rm drag}=
\rho_{hg} S_{cl} (v_{hg} -V) |(v_{hg} -V)|$,
acting to reduce the cloud acceleration.
   But if the cloud acceleration $a_{cl}$
decreases, the covering factor of the clouds (depending
on acceleration, see above) can become large enough for
the absorption in the strong lines to increase.
  Consequently, the radiation pressure force falls due to
this absorption.
   This leads to a nonlinear change of
acceleration, which can became negative.
   As the velocity of the hot gas increases further, it becomes
higher than the cloud's velocity at some point.
  Then the drag force changes sign and it acts to accelerate the clouds.
   This increases the radiation pressure force
again, and the process repeats.
   The velocity dependence
$V(r)$ in this case looks like a ``ladder'' with segments of
nearly constant velocity.
  Figure 3 shows the velocity profile for an illustrative case.
   This results in the appearance
of repeated narrow absorption details in the spectrum.
But in this case stability of the narrow details means
stability of the velocity (and acceleration) structure,
and not absence of acceleration in some increased
density regions of the flow as usually assumed.

We used our dynamical model to calculate the hot and
cold phase gas dynamics.

            Figure 4 shows the resulting
calculated spectrum and the
observed one for the quasar Q1303+308
(Vilkoviskij and Irwin 2001).  One can
see that there is rather good agreement
in many details of the calculated
and observed spectra.
   This supports the dynamical model and the
above-described mechanism of ``line-locking''.

         Figure 5 shows in more detail that even some fine
features of the calculated spectrum are
similar to those observed in one of
the CIV absorption throughs.
Of course, there are also differences but the
agreement is satisfactory.

     The main parameters of the model are:
The hot gas mass flow rate, $F=0.1 M_\odot/{\rm yr}$;
the black hole mass, $M_{bh}=1.3\times10^8M_\odot$;
the total mass of the compact stellar cluster,
$M_{csc}=2.5\times109 M_\odot$;
 the power in the black-body
component, $L_{p}=1.5 \times 1.5\times 10^{47}$ erg/s
at the temperature  $T_p=2.1\times 104$ K;
the energy flux in the power-law continuum,
$L_{\rm cont}= 10^{46}$ erg/s which is assumed
to have a slope of $1.7$ (i.e., $F_\nu \sim \nu^{-1.7}$).
The mass of a single cold cloud is $5\times 10^{-16}M_\odot$.

        The mass-flow rate contained in the small cold
clouds was modeled with a linear interpolation between
the following values.

\begin{tabular}{ | c | c | c |c | c |}
\hline
$r(pc)$ & $10^{-3}$ & $2$ & $10$ & $20$ \\
\hline
mass flux($M_\odot$/{\rm yr}) & $10^{-6}$ & $0.005$ & $0.6$ & $0.3$\\
\hline
\end{tabular}

        The calculated spectrum of the same object in
both UV and X-ray bands is shown in Figure 6. One can see
that our model predicts the ``X-ray quiet'' soft X-ray
spectrum as observed for BAL QSOs  (Brandt, Laor, Wills
2000; Green et al. 2001).

  Our present model  does not
include  emission and absorption lines in the X-ray band.
 Inclusion of these is planned in the next version of
the model.
 The spectra of AGNs are poorly
known in the border of the UV and X-ray bands,
where the so-called ``soft excess''
is present in many objects. We note that in our model the
intensity of the emission lines are relatively strong
relative to the continuum
in the region of several tens of  eV.
  Thus these lines can partly ``mimic'' the
soft excess as discussed by Crummy et al. (2006).

\section{Conclusions}

As remarked by Crenshaw, Kraemer, and George (2003)
``The study of intrinsic absorption is still in infancy,
and detailed comparison between the observations and
models has just begun''. Of course, the dynamical model
which we used to calculate the Q1303+308 spectrum,
involves some simplifications. Nevertheless, it is the
only present-day model, which allows calculation of the
spectrum with ``line-locking''.  The model is of course being
developed further.

The mass outflow rate is a fundamental aspect
of an AGN and it must be in accordance with the
classical unification scheme of Antonucci (1993), which
is strongly supported by observations.
  But a model of an
actual AGN has to take into account not only the
geometrical unification, but also the evolutionary stage
of the AGN, the type of its host galaxy, the orientation
of the accretion disk relative to the galactic disk
plane, and other details.
  In the unified model of AGN
mass outflow, the deep absorptions in the UV lines is
supposed to be seen in the transition angles between the
types AGN1 and AGN2.
   The strongest UV absorptions in
Seyferts are indeed visible in the spectra of the
``intermediate'' types of Seyferts Sy1.5 tp Sy1.9.
But more statistical investigations are needed to prove this
rule.

    The BAL QSOs present about $15\%$ of radio-quiet QSOs,
but the ``intrinsic'' fraction of BAL QSO, taking into
account the K-correction, is up to $22\%$ (Hewett and Foltz
2003).  A large part ($\sim 50\%$) of Seyfert galaxies showing
both absorption lines in UV and/or warm absorbers can
be explained by larger covering factors.
  This is because
of the smaller velocities and the velocity gradients.
Partly it may reflect the larger number of high-quality
spectra for the closest Seyferts. The fraction of absorbed
Seyfert galaxies was estimated as $\sim 0.1$ when investigated by IUE
satellite with good spectral resolution.

   It is possible that the internal surface of the
obscuring torus is not the only source of the outflowing
matter of the cool clouds.
  For example, the source could  be the red giant
winds and/or gas resulting from stellar encounters of the
compact stellar cluster.
   Also, the absorbing gas may be present in the
hollow central region of the OT due to mass outflow from the
accretion disk surface.
   This may occur if the disk is
geometrically thick in the outer region, closer to the
OT due to gravitational instability and/or magnetic
fields.
   The hydromagnetic winds from accretion disks
have been studied in detail by Ustyugova et al. (1999),
Romanova et al. (2004), Romanova et al. (2005), and
Ustyugova et al. (2006).  Hydromagnetic models for the
emission and absorption properties of NGC 5548 have been
developed by Bottorff et al. (1997) and Battorff,
Korista and Shlosman (2000).

  The above-described unified dynamical model
explains many details of the complicated spectra observed in
different BALQSOs.  This argues for the validity of
the model for the BALQSOs.

       This work was supported by the CRDF
grant KP2-2555-AL-03.

\begin{figure*}[t]
\centerline{\epsfig{file=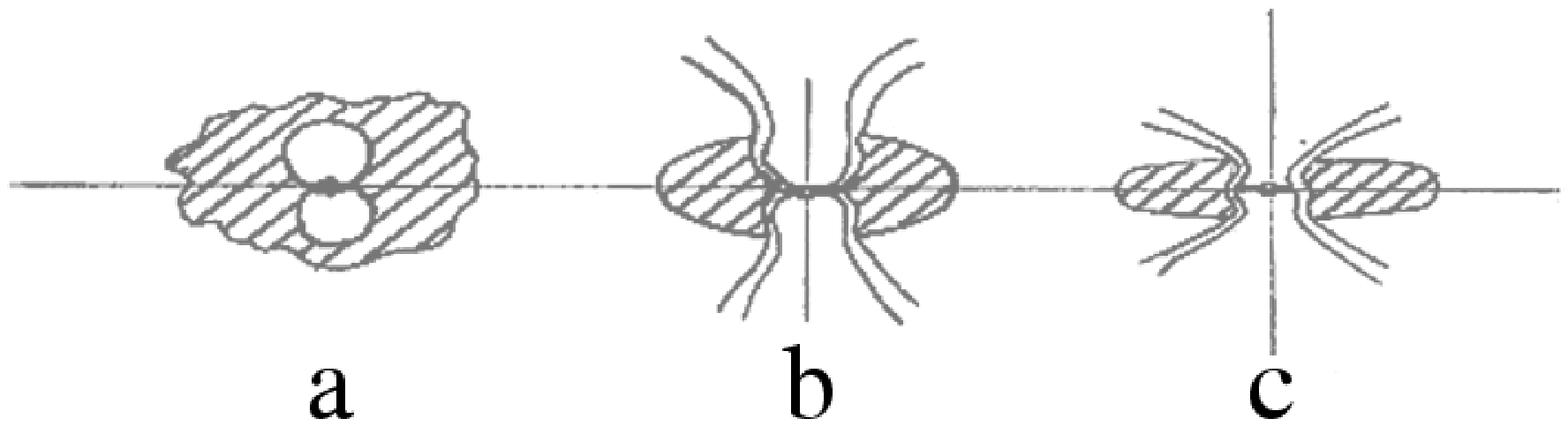}}
\caption{Sketches of the three stages of an AGN's duty cycle.}
\end{figure*}

\begin{figure*}[t]
\centerline{\epsfig{file=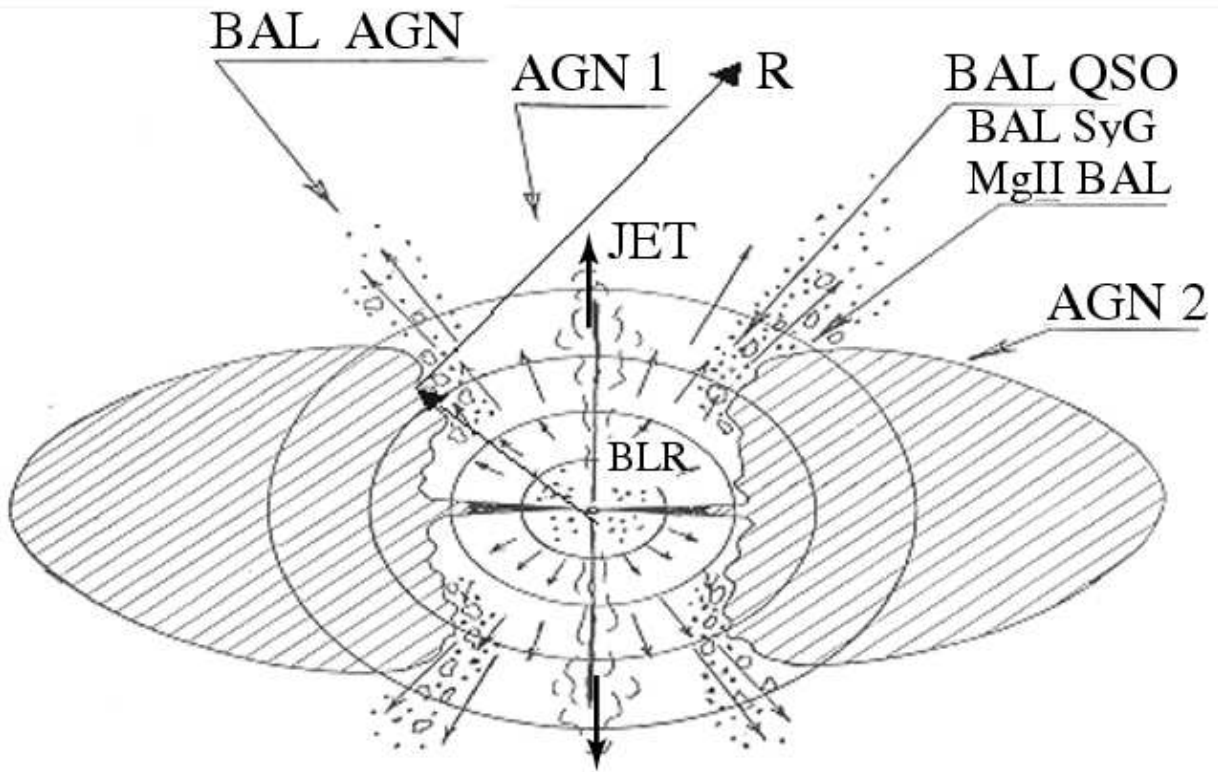}}
\caption{Sketch of unified AGN model (V99).}
\end{figure*}

\begin{figure*}[t]
\centerline{\epsfig{file=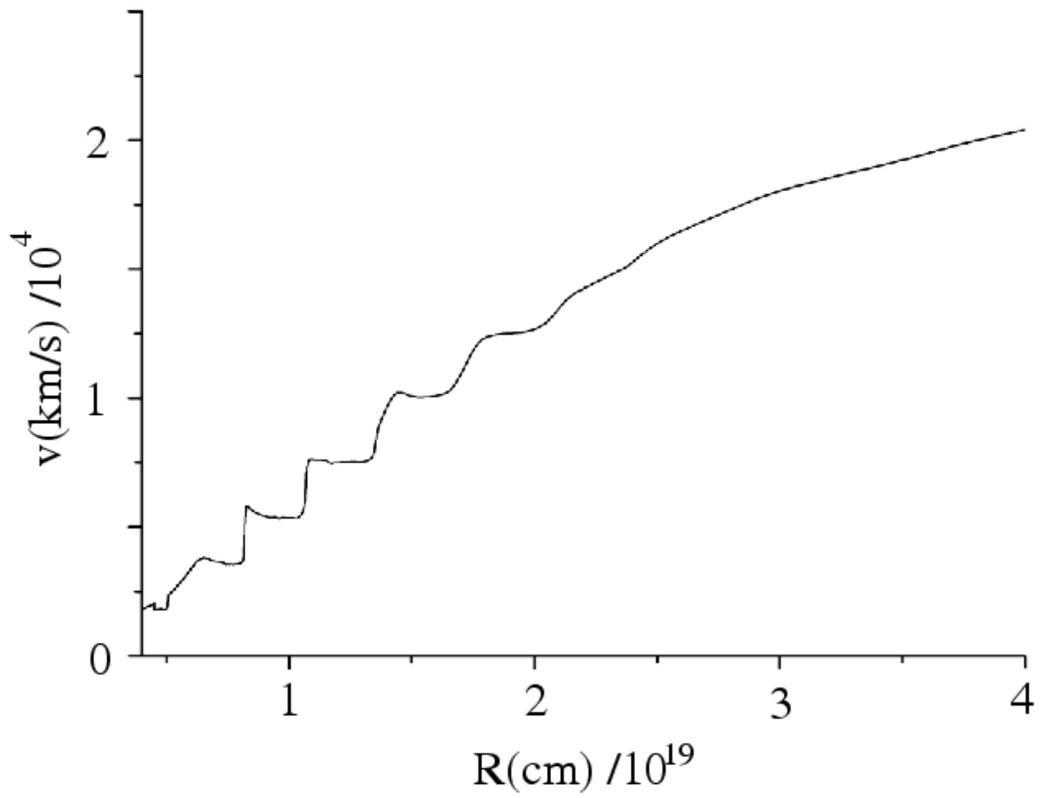}}
\caption{Dependence of the cold cloud velocity on the distance
from the central object for the case discussed in \S 5.}
\end{figure*}

\begin{figure*}[t]
\centerline{\epsfig{file=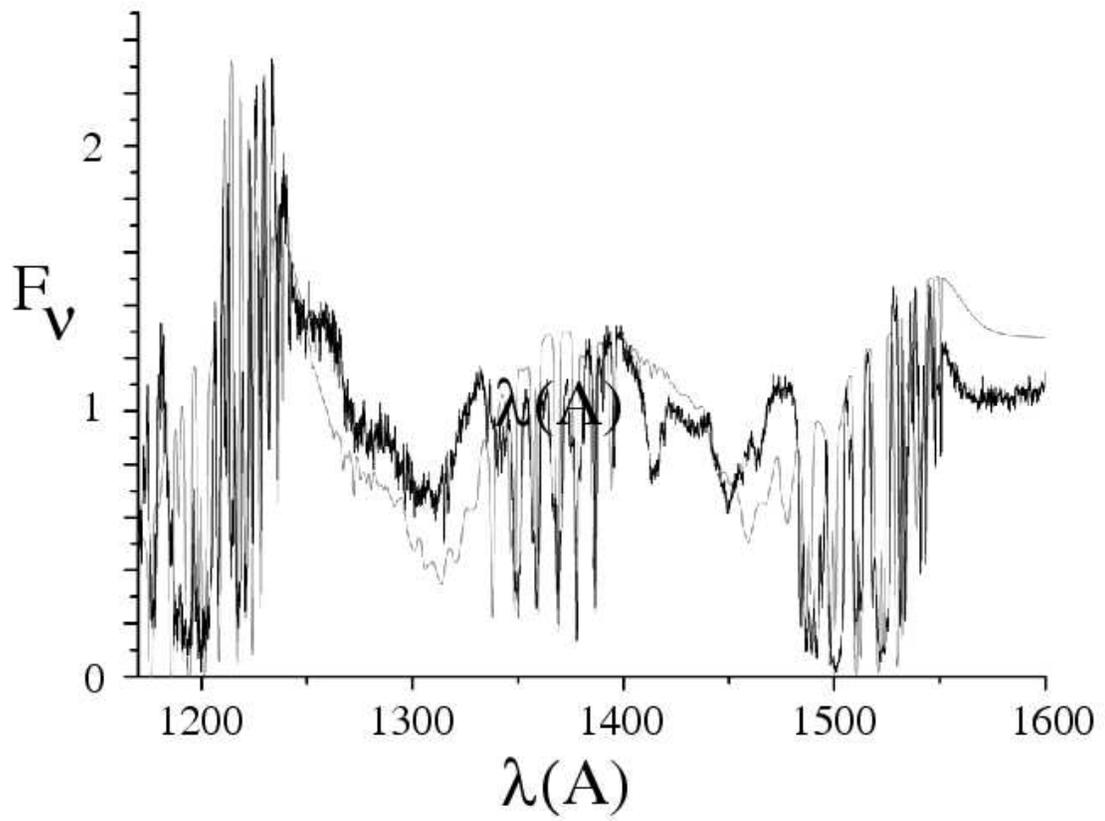}}
\caption{The observed (black) and calculated (gray) spectra
of the quasar q1303+308.}
\end{figure*}

\begin{figure*}[t]
\centerline{\epsfig{file=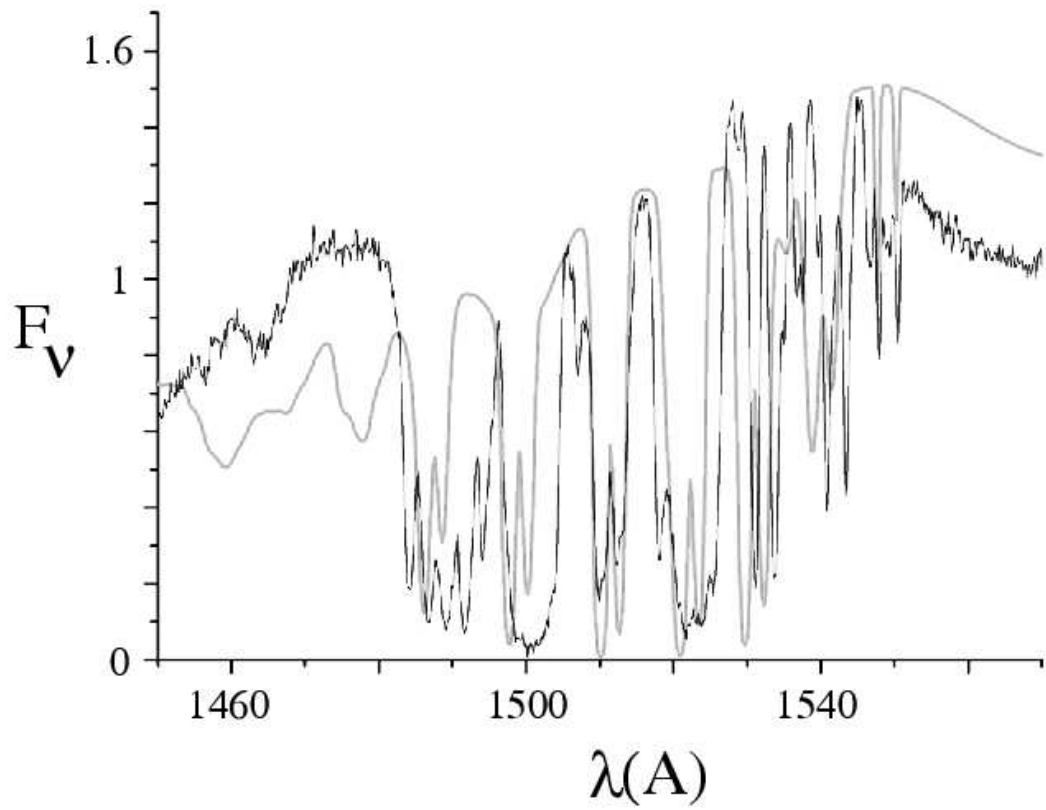}}
\caption{The observed (black) and calculated (gray) spectra
of the quasar q1303+308 at higer resolution showing the
CIV lines.}
\end{figure*}

\begin{figure*}[t]
\centerline{\epsfig{file=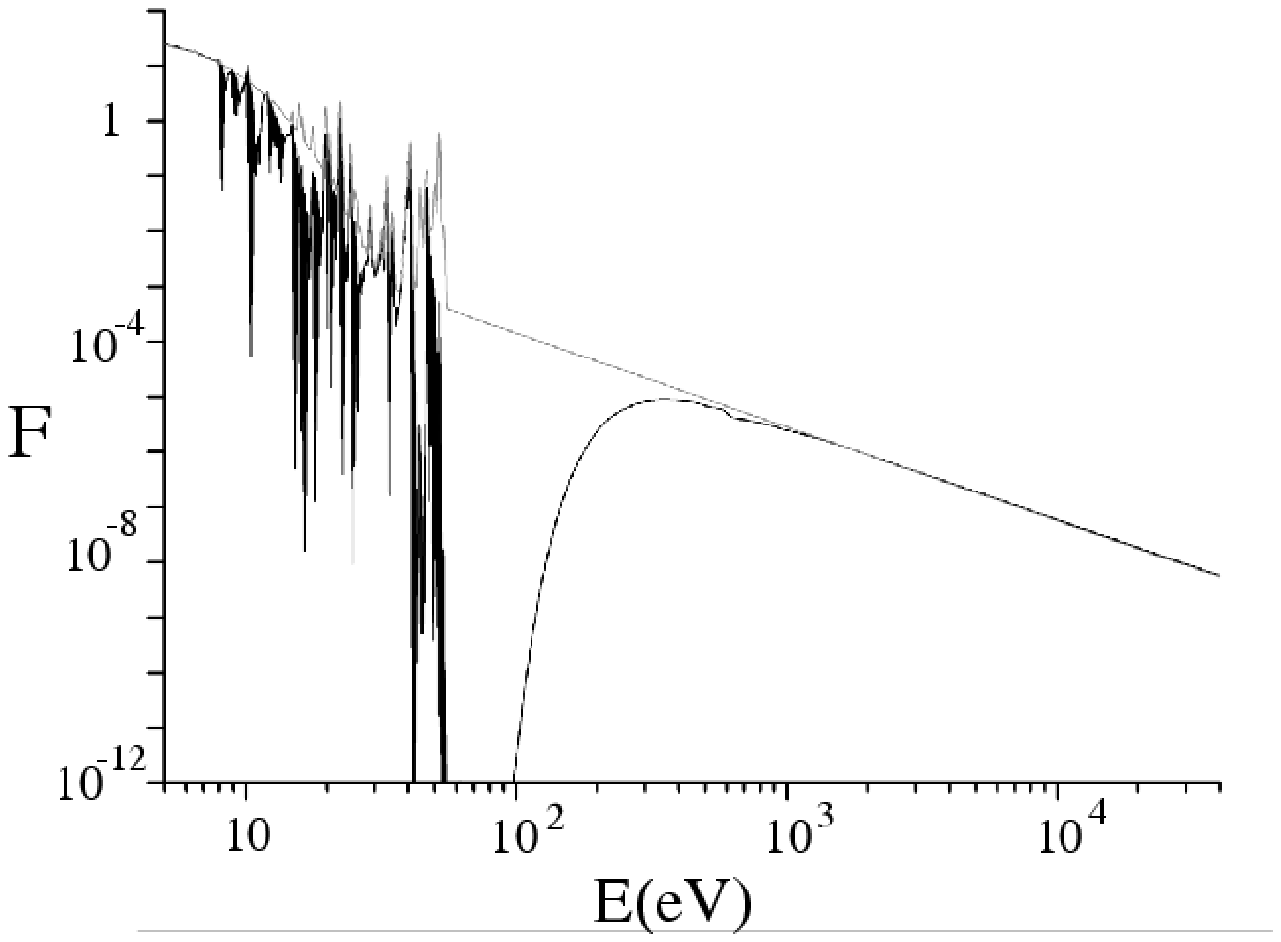}}
\caption{Predicted photon spectrum before (gray) and after (black)
absorption.}
\end{figure*}


\begin{thebibliography}{10}

\bibitem{}Antonucci, R. 1993, ARA\&A, 31, 473

\bibitem{}Antonucci, R.R.J., \& Miller, J.S. 1985, ApJ 297, 621

\bibitem{} Arav, N. \& Begelman, M. 1994, ApJ 434, 479

\bibitem{} Arav, N., Korista, K. T., \& de Kool, M. 2002,
ApJ, 566, 699

\bibitem{}Arav, N., Kaastra, J., Steenbrugge, K.,
Brinkman, B., Edelson, R., Korista, K., \& de Kool, M.
 2003 ApJ 590, 174

\bibitem{}Barlow, T.A., \& Sargent, W.L.W. 1997, AJ, 113, 136

\bibitem{}Begelman, M.C., de Kool, M., \& Sikora, M. 1991, ApJ 382, 416

\bibitem{}Bianchi, S., \& Matt, G. 2002, A\&A, 387, 76

\bibitem{}Bianchi, S., Matt, G., Balestra, I., Guainazzi, M.,
\& Perola, G. C. 2004, A\&A, 422, 65

\bibitem{}Bianchi, S., Matt, G., Nicastro, F., Porquet, D.,
\& Dubau, J. 2005, MNRAS
357, 599

\bibitem{}Bisnovatyi-Kogan, G.S.,
\& Lovelace, R.V.E. 2001, New Astron. Rev., 45, 663

\bibitem{}Blandford, R.D., \& Znajek, R.L. 1977, MNRAS,179, 433

\bibitem{}Blustin, A.J., Page, M.J., Fuerst, V., Branduardi-Raymont, G.,
\& Ashton, E.C. 2005, A\&A 431,
  111

\bibitem{}Boeker, T, Sarzi, M., McLaughlin, D.E.,
van der Marel, R.P., Rix, H-W., Ho, L.C., \& Shields, J.C.
  2004, AJ 127, 105

\bibitem{}Bottorff, M., Korista, K.T., Shlosman, I.,
\& Blandford, R.D. 1997, ApJ 479, 200

\bibitem{}Bottorff, M., Korista, K.T., \& Shlosman, I. 2000, ApJ 537, 134

\bibitem{}Brandt, W.N., Laor, A., \& Wills, B.J. 2000, ApJ 528, 637

\bibitem{}Burbidge, E.M. 1970, ApJ 160, L33

\bibitem{}Castor, J., Abbot, D., \& Klein, R. 1975, ApJ 195, 157

\bibitem{}Crummy, J., Fabian, A.C., Gallo, L.,
\ Ros, R.R. 2006, MNRAS, 365, 1067

\bibitem{}Crenshaw, D.M. 1997, in ASP Conf. Series, Vol. 113,
``Emission lines in Active Galactics: New
  Methods and Techniques,'' eds. B.M. Peterson, F.-Z. Cheng
\& A.S. Wilson, IAU Colloq.159, p.240

\bibitem{}Crenshaw, D.M., Maran, S. P.,
\& Mushotzky, R. F. 1998, ApJ, 496, 749

\bibitem{}Crenshaw D.M., Kraemer S.B., Bogges A.,
Stephen P., Mushotzky R.F., \& Wu C-C., 1999,
 ApJ, 516, 750

\bibitem{}Crenshaw, D. M., Kraemer, S. B.,
\& George, I. M. 2003, ARA\&A, 41, 117

\bibitem{}Czerny, B.,\& Lehto, H.J., 1997, MNRAS 285, 365

\bibitem{}Czerny, B., Schvarzenberg-Czerny, A.,
\& Loska, Z., 1999, MNRA 303, 148

\bibitem{}Davies, R. I., Tacconi, L. J., \& Genzel, R. 2004a, ApJ, 602, 148

\bibitem{}Davies, R. I., Tacconi, L. J.,\& Genzel, R., 2004b, ApJ, 613, 681

\bibitem{}deKool, M., Korista, K.T., \& Arav, N., 2002, ApJ, 580, 54

\bibitem{}Elvis, M., 2000, ApJ, 545, 63

\bibitem{}Fabian, A.C., Guilbert, P.W., Arnaud, K.A.,
Shafer, R.A., Tennant, A.F., \& Ward, M.J.
   1986, MNRAS  218, 457

\bibitem{}Foltz, C.B., Weymann, R.J., Peterson, B.M.,
Sun, L., Malkan, M.A., \& Chaffee, F.H. 1986, ApJ,
  307, 504

\bibitem{}Ganguly, R., Bond, N.A., Charlton, J.C.,
Eracleous, M., Brandt, W.N. \& Churchill, C.W. 2001,
  ApJ 549, 133

\bibitem{}George, I. M., Turner, T. J., Mushotzky, R.,
Nandra, K., \& Netzer, H., 1998a, ApJ, 503, 174

\bibitem{}George, I. M., Turner, T. J., Netzer, H.,
Nandra, K., Mushotzky, R., \& Yaqoob, T. 1998b, ApJS,
  114, 73

\bibitem{}Green, P.J., Aldcroft, T.L., Mathur, S.M.,
Wilkes, B.J., \& Elvis, M. 2001, ApJ 558, 109

\bibitem{}Haas, M., Klaas, U., Müller, S.A.H., Bertoldi, F., Camenzind, M.,
Chini, R., Krause, O., Lemke, D., Meisenheimer, K.,
Richards, P.J.,\& Wilkes, B.J. 2003, A\&A 402, 87

\bibitem{}Hewett, P.C., \& Foltz, C.B. 2003, ApJ 125, 1784

\bibitem{}Hines, D.C., \& Wills, B.J. 1995, ApJ 448, L69

\bibitem{}Hines, D.C., Schmidt, G. D., Gordon, K, D.,
Smith, P. S., Wills, B. J.,
Allen, R. G., \& Sitko, M. L. 2001,
ApJ, 563, 512

\bibitem{}Kaastra, J. S., Mewe, R., Liedahl, D. A.,
Komossa, S., \& Brinkman, A. C. 2000, A\&A 354, L83

\bibitem{}Kaspi, S., Brandt, W.N., Netzer, H., Sambruna, R.,
Chartas, G., Garmire, G.P., \& Nousek, J.A. 2000,
 ApJ 535, L17

\bibitem{}Kaspi, S., Brandt, W.N., George, I. M.,
Netzer H., et al.,
Kaspi, S., Brandt, W. N., George, I. M., Netzer, H., Crenshaw, D. M.,
Gabel, J. R., Hamann, F. W., Kaiser, M. E., Koratkar, A.,
Kraemer, S. B., Kriss, G. A., Mathur, S., Mushotzky, R. F.,
Nandra, K., Peterson, B. M., Shields, J. C., Turner, T. J., \& Zheng, W.
2002, ApJ, 574, 643

\bibitem{}Kriss, G. A., Davidsen, A. F., Blair, W. P.,
Bowers, C. W., Dixon, W. V., Durrance, S. T.,
 Feldman, P. D., Ferguson, H. C., Henry, R. C., Kimble, R. A., Kruk, J. W.,
Long, Knox S., Moos, H. W., \& Vancura, O.
 1992, ApJ, 392, 485

\bibitem{}Kriss, G. A., Davidsen, A. F., Zheng, W.,
Kruk, J. W., \& Espey, B. R. 1995, ApJ, 454, L7

\bibitem{}Kriss, G. A., Espey, B. R., Krolik, J. H.,
Tsvetanov, Z., Zheng, W., \& Davidsen, A.F. 1996, ApJ,
 467, 622

\bibitem{}Kawaguchi, T., Shimura, T., \& Mineshige, S. 2001, ApJ, 546, 966

\bibitem{}Komossa, S., \& Hasinger, G. 2003, in:
``XEUS - Studying the Evolution of the Hot Universe,'' G.
 Hasinger et al. (eds), MPE Report 281, 85

\bibitem{}Kormendy, J., 2001, RevMexAA 10, 69

\bibitem{}Krolick, J.H., McKee, C.F.,
\& Tarter, C.B. 1981, ApJ 249, 422

\bibitem{}Krolick, J. H., \& Begelman, M. C. 1988, ApJ 329, 702

\bibitem{}Krolick, J. H., \& Kriss, G. A. 1995, ApJ, 447, 512

\bibitem{}Krolick, J.H., \& Kriss, G. A. 2001, ApJ, 561, 684

\bibitem{}Kuznetsov, O.A., Lovelace, R.V.E., Romanova, M.M., \&
Chechetkin, V.M. 1999, ApJ, 514, 691

\bibitem{}Kwan, J., \& Krolik, J.N. 1981,  ApJ 250, 478

\bibitem{}Kwan, J. 1990, ApJ 353,121

\bibitem{}Lynden-Bell, D. 1996, MNRAS, 279, 389

\bibitem{}Lovelace, R.V.E. 1976, Nature 262, 619

\bibitem{}Lovelace, R.V.E., Romanova, M.M.,
\& Biermann, P.L. 1998, A\&A, 338, 856

\bibitem{}Lovelace,R.V.E., \& Romanova, M.M. 2003, ApJ 596, L159

\bibitem{}Lucy, L.B., \& Solomon, R.P. 1970, ApJ 159, 879

\bibitem{}Lynds, C.R. 1967, ApJ 147, 396

\bibitem{}Matt, G., Bianchi, S., D'Ammando, F.,
\& Martoccia, A. 2004a, A\&A, 421, 473

\bibitem{}Matt, G., Bianchi, S., Guainazzi, M.,
\& Molendi, S. 2004b, A\&A 414, 155

\bibitem{}Mency, N., Cavalierre, A., Fontana, A., Giallongo, E., Poli, F.,
\& Vittoini, V. 2003,
 ApJ 587, L63

\bibitem{}Murray, N., Chang, J., Grossmann, S.A.,
\& Voit, G.M. 1995, ApJ 451, 498

\bibitem{}Murray, N., \& Chang, J. 1995, ApJ 454, 105

\bibitem{}Murray, N., \& Chang, J. 1998, ApJ 494, 125

\bibitem{}
    Netzer, H., Kaspi, S., Behar, E., Brandt, W. N., Chelouche, D.,
George, I. M., Crenshaw, D. M., Gabel, J. R., Hamann, F. W.,
Kraemer, S. B., Kriss, G. A.,  Nandra, K., Peterson, B. M.,
Shields, J. C., \& Turner, T. J.
2003, ApJ 599, 933

\bibitem{}Ogle, P.M. 1997, ASPC, 128, 78

\bibitem{}Ogle, P. M., Cohen, M. H., Muller, J. S., Tran, H. D.,
Goodrich, R. W., \& Martel, A.R., 1999, ApJ
  Suppl, 125,1

\bibitem{}Ogle, P.M., Brookings, T., Canizares, C.R., Lee, J.C.,
\& Marshall, H.L. 2003, A\&A 402, 849

\bibitem{}Ogle, P.M., Marshall, H.D., Lee, J.C.,
\& Canizaros, C.R. 2000, ApJ 545, L81

\bibitem{}Proga, D., Stone, J. M., \& Kallman, T. R. 2000, ApJ, 543, 686

\bibitem{}Reeves, J. N., O'Brein, P. T., \& Ward, M. J. 2003,  ApJ 593, L65

\bibitem{}Reeves, J. N., Nandra, K., George, I. M., Pounds, K. A.,
Turner, T. J., \& Yaqoob, T. 2004, ApJ 602,
  648

\bibitem{}Reynolds, C. S. 1997, MNRAS, 286, 513

\bibitem{}Romanova, M.M., \& Lovelace, R.V.E. 1997, ApJ, 475, 97

\bibitem{}Romanova, M.M., Ustyugova, G.V.,  Koldoba, A.V.,
\& Lovelace, R.V.E. 2005, ApJ, 635, L165

\bibitem{}Romanova, M.M., Ustyugova, G.V.,  Koldoba, A.V.,
\& Lovelace, R.V.E. 2004, ApJ, 616, 151

\bibitem{}Romanova, M.M., Ustyugova, G.V., Koldoba, A.V.,
\& Lovelace, R.V.E. 2005, ApJ, 635, L165

\bibitem{}Rubin, V.C., Graham, J.A., \& Kenney, J.D.P. 1992,
ApJ, 394, L9



\bibitem{}Sanders, D.B., Soifer, B.T., Elias, J.H.,
Neugebauer, G., \& Mathews, K. 1988, ApJ 328,
    L 35

\bibitem{}Sazonov, S.Yu., Ostriker, J.P., \& Sunyaev, R.A. 2004,
MNRAS, 347, 144

\bibitem{}Syer, D., Clarke, C. J., \& Rees, M. J. 1991, MNRAS 250, 505

\bibitem{}Turnshek, D.A. 1987, in ``QSO absorption lines:
Probing the Universe'' (STScl publication), eds.
  Blades, Normann, \& Turnshek, 2, p.17

\bibitem{}Ustyugova, G.V., Koldoba, A. V.,
Romanova, M.M., Chechetkin, V.M., \& Lovelace, R.V.E.
   1999, ApJ 516, 221

\bibitem{}Ustyugova, G.V.,  Koldoba, A. V.,
Romanova, M.M., \& Lovelace, R.V.E.
   2006, ApJ, in press

\bibitem{}Veron-Cetty, M.P., \& Veron, P. 2000, Astron. Astrophys. Rev .10, 81

\bibitem{}Vilkoviskij, E. Y., \& Nosov, I. V. 1994,
in ``QSO Absorption Lines'' ed. by Meylan G., p.241

\bibitem{}Vilkoviskij, E. Y., Karpova, O. G.,
\& Nosov, I. V. 1996, Astron. Rep. 40, 305

\bibitem{}Vilkoviskij, E. Y., \& Karpova, O. G. 1996, Astron. Lett. 22, 148

\bibitem{}Vilkoviskij, E. Y., Efimov, S. N.,
Karpova, O. G., \& Pavlova, L. A. 1999, MNRAS
  309, 80

\bibitem{}Vilkoviskij, E.Y., \& Irwin, M.J. 2001, MNRAS, 321, 4

\bibitem{}Vilkoviskij, E. Y., \& Czerny B. 2002, A\&A 387, 804

\bibitem{}Wada, K., Norman, C.A., \& Colin, A. 2002, ApJ, 566, L21

\bibitem{}Wada, K., Norman, C.A., \& Colin, A. 2001, ApJ, 547, 172

\bibitem{}Weymann, R. J.,  Carswell, R. F.,
\& Smith, M. G. 1981, ARA\&A 19, 41

\bibitem{}Weymann, R.J., Scott, J.S., Shiano, A.V.R.,
\& Christiansen, W.A. 1982, ApJ 262, 497

\bibitem{}Weymann, R. J.,  Morris, S.L., Foltz, C.B.,
\& Hewett,  P. C. 1991, ApJ 373, 23

\bibitem{}Weymann, R.J., Morris, S.L., Gray, M.E.,
\& Hutchings, J.B. 1997, ApJ 483, 717

\bibitem{}Weymann ,R.J. 2002, in:
``Mass Outflow in AGN: New Perspectives,'' eds. D.Crenshaw,
     S. Kraemer, \& I. George,  ASP Conf. Series, Vol. 255, p. 29

\end{thebibliography}
\end{document}